\documentclass[pdflatex,sn-mathphys-num]{sn-jnl}

\usepackage{graphicx}%
\usepackage{multirow}%
\usepackage{amsmath,amssymb,amsfonts}%
\usepackage{amsthm}%
\usepackage{mathrsfs}%
\usepackage[title]{appendix}%
\usepackage{xcolor}%
\usepackage{textcomp}%
\usepackage{manyfoot}%
\usepackage{booktabs}%
\usepackage{algorithm}%
\usepackage{algorithmicx}%
\usepackage{algpseudocode}%
\usepackage{listings}%

\usepackage{subcaption}
\usepackage{tabularx}
\usepackage{float}

\theoremstyle{thmstyleone}%
\theoremstyle{thmstyletwo}%

\theoremstyle{thmstylethree}%

\begin{document}

\title[Article Title]{The software space of science}


\author[1]{\fnm{Zhouming} \sur{Wu}}\email{wu.zhoum@northeastern.edu}

\author*[2]{\fnm{Dakota} \sur{Murray}}\email{dsmurray@albany.edu}


\affil[1]{\orgdiv{Network Science Institute}, \orgname{Northeastern University}, \orgaddress{\street{177 Huntington Ave}, \city{Boston}, \postcode{02115}, \state{MA}, \country{USA}}}

\affil[2]{\orgdiv{College of Emergency Preparedness, Homeland Security and Cybersecurity}, \orgname{University at Albany, State University of New York}, \orgaddress{\street{1220 Washington Avenue}, \city{Albany}, \postcode{12226}, \state{NY}, \country{USA}}}

\abstract{
Science advances not only through the accumulation of facts but also through the evolution of tools. 
Crucially, tools are rarely used in isolation.
They form tool portfolios, combinations shaped by a discipline's workflows and analytical demands. 
Software, near-ubiquitous in modern research and traceable across the published literature, offers a unique window to study tool use in science.
Here, we map the software space of science by analyzing mentions to software from 1.3 million publications from 2004 to 2021. 
We construct a network of 520 software tools linked by disciplinary co-usage, with link strength weighted by proximity based on revealed comparative advantage.
This network reveals a structured landscape in which tools cluster into 8 functional communities, including \textit{computing and statistics}, \textit{wet lab instrumentation}, and several bioinformatics specializations, with
each discipline occupying a distinct position reflecting its characteristic tool portfolios. 
The breadth of a discipline's tool portfolio is shaped by the nature of its research workflow: fields combining experimental and computational tasks draw on multiple communities, while those with narrower methodological demands concentrate in one. These structural differences are stable across the observation period. 
At the same time, across all broad disciplinary categories, disciplinary tool portfolios are crystallizing, settling on a common set of tools.
}




\maketitle

\section{Introduction}\label{sec1}
Science advances not only through the accumulation of facts but also through the evolution of tools~\cite{dyson1999sun}. 
X-ray crystallography revealed the structure of DNA, while the polymerase chain reaction made modern molecular biology possible. 
More recently, deep learning frameworks combined with tools like \texttt{AlphaFold} have achieved breakthroughs in protein structure prediction that decades of traditional methods could not~\cite{jumper2021highly, tunyasuvunakool2021highly, bertoline2023before}. 
This pattern generalizes across science; a study of Nobel Prize discoveries found that the  majority relied on sophisticated methods or instruments, ranging from statistical techniques to particle accelerators~\cite{krauss2024redefining}. 

Despite their apparent importance to scientific progress, we know remarkably little about how tools are adopted, spread, and evolve across science. 
The evolution of ideas and social structures in science can be investigated via citation traces~\cite{small1973co, radicchi2011citation} and co-authorship~\cite{wuchty2007increasing, borner2012design}.
Yet unlike citation linkages or co-authorship, tools are not reliably indexed in bibliometric data, challenging systematic and large-scale investigation.

Among the many types of scientific tools, software occupies a unique position: it is pervasive across nearly all disciplines and, crucially, it leaves traces in the published literature that can be systematically identified and analyzed.
Physical instruments, even when named in a paper's methodology, are reported inconsistently across papers, challenging large-scale extraction and disambiguation. 
Software tools, by contrast, have distinctive names that are more easily detected and disambiguated in text~\cite{howison2016software} and advances in text mining have made identification possible at scale, from curated corpora~\cite{du2021softcite, du2022understanding} to millions of articles~\cite{istrate2022large}.

Beyond their tractability, software also sheds light on a fundamental aspect of tool use in science: tools are rarely used in isolation. 
For example, in computationally intensive fields, a single analysis may require chaining together multiple tools, a practice now formalized through workflow management systems~\cite{wratten2021reproducible}. 
More broadly, tools often depend on shared data formats, compatible interfaces, or overlapping user communities, so that the adoption of one tool constrains or encourages the adoption of others. 
Tool adoption is therefore best understood not at the level of individual scientists choosing individual tools, but at the level of scientific fields collectively developing a \textit{tool portfolio} --- the set of tools routinely combined in common disciplinary workflows.

Building on this insight, we investigate the tool portfolios of scientific fields by constructing a ``software space'' in which tools are linked based on patterns of disciplinary co-adoption, on the premise that tools serving related functions tend to appear together in the same fields. 
This approach draws inspiration from the ``product space" in economics~\cite{hidalgo2007product}, which maps relationships among products based on countries' co-export patterns.
Applied to scientific software, this map reveals distinct communities of tools, characterizes a discipline's tool portfolio, and can be used to trace the evolution of disciplinary tool use over time. 

\section{Background}\label{sec2}

\subsection{Software as a scientific tool}
We define ``tools" broadly as the artifacts, whether physical, procedural, or computational, that are used as part of a workflow to conduct experiments, collect or process data, conduct analysis, and so on.
This definition encompasses laboratory instruments, standardized protocols, statistical methods, and software applications. 
Tools are distinguished from theories or findings by their functional character: they enable the production of scientific knowledge rather than constituting that knowledge directly.

Software occupies a distinctive position within this broader category. 
We define software as any computational tool that researchers use to collect, process, or analyze data. This includes statistical packages (\texttt{SPSS}, \texttt{Stata}), programming languages and their libraries (\texttt{Python}, \texttt{NumPy}, \texttt{TensorFlow}), domain-specific applications (\texttt{ImageJ}, \texttt{BLAST}), and general purpose tools for data handling and visualization (\texttt{Excel}, \texttt{SQL}, \texttt{GraphPad Prism}), among others. This definition is deliberately broad, as the boundary between a ``tool" and a ``method" is often blurred in practice. 
\texttt{BLAST}, for instance, is both a sequence alignment algorithm and a software tool that implements it. 
Rather than attempting to draw this boundary, we focus on named software entities that researchers explicitly mention in publications, which can be identified and tracked regardless of how one classifies them.

Software also has several qualities that distinguish it from other types of tools. 
First, it is uniquely ubiquitous: whereas particular types of scientific equipment may be confined to a single discipline, software is used across a wide range of fields, and in computationally intensive disciplines it has become the primary medium through which methods are implemented and shared~\cite{hocquet2024software}. 
Second, software is uniquely portable and accessible. 
Physical tools can be expensive, slow to procure, and require specialized training; much of the software used in science, by contrast, can be obtained at low cost or for free and deployed with relatively little specialized knowledge. 
In this way, software can make even specialized techniques widely accessible, as when deep learning frameworks like \texttt{PyTorch}~\cite{paszke2019pytorch} lower the barrier to neural network training. 
This reduces the external frictions, such as cost, access, and infrastructure, that may confound studies of tool adoption, making software a cleaner signal of disciplinary practice and preference. 
Finally, software is uniquely visible in the scientific literature relative to other tools.
Physical instruments are typically described in ways that vary widely across papers, with inconsistent reporting of manufacturer names, model numbers, and configuration details.
This makes large-scale extraction and disambiguation difficult. 
Software, by contrast, tends to have distinctive names that are more easily detected and disambiguated in unstructured text~\cite{howison2016software}. A tool like \texttt{DESeq2} or \texttt{ImageJ} is referred to by name across thousands of papers in ways that a centrifuge or mass spectrometer rarely is, making software mentions a tractable signal of tool use even in the absence of formal citation.
Together, these qualities facilitate large-scale analysis of scientific software.

\subsection{Tracking scientific software}
Software is more visible in the literature than other tools, though identifying and extracting those traces reliably at scale remains an open challenge.
Two directions have emerged in response to this challenge: one seeking to change how software is cited, and one seeking to work around the limitations of existing citation practice.

The first direction could open up software to large-scale bibliometric analysis by reforming how it is published and cited. 
If software were treated as a first-class research object---with stable identifiers, versioned releases, and consistent citation practices---it would become as traceable as any other scholarly output. 
Initiatives like the \textit{FAIR} principles advocate for exactly this~\cite{barker2022fair}, and repositories like Zenodo and Figshare now facilitate software indexing in ways that make it citable like any other research output~\cite{patel2023fair}. 
A related proposal is the ``software paper'' --- a publication written describing a piece of software that serves as a citable proxy for the tool itself~\cite{romano2020software-paper}; venues such as the \textit{Journal of Open Source Software} exist specifically to publish such works. 
Yet significant challenges remain. 
Software evolves over time, complicating citation to specific versions, and efforts to standardize citation norms~\cite{smith2016citation} have not taken hold uniformly across disciplines~\cite{wang2024software, li2019lme4, pan2016disciplinary}. 
GitHub exemplifies this.
As arguably the largest repository of scientific software~\cite{escamilla2022github, trujillo2022penumbra}, there remains a lack of common practices for citing repositories or referencing specific versions, and URLs remain the default. 
These normative efforts remain ongoing and have improved software credit allocation at the margins, but have yet to produce the consistency required for large-scale bibliometric analysis. 

The second direction aims to identify software use automatically from unstructured content in the published literature, commonly referred to as ``software mentions''. 
Early approaches relied on curated lists of easily identifiable tools~\cite{li2019lme4, pan2018bibliometric, li2017R-software}. 
More recent work leverages advances in text mining to detect software mentions in unstructured full text at scale, an approach that treats any mention of a named software entity as a signal of use, without requiring formal citations or software papers (though these may be incorporated).
The growing availability of full-text corpora, combined with improvements in machine learning for information extraction, has enabled the construction of large-scale empirical datasets of software mentions~\cite{du2021softcite, du2022understanding, istrate2022large}. 
While such approaches have improved greatly over time, they also often produce noisy results, struggle to identify clear instances of actual use (rather than mere mention), and have difficulty disambiguating software entities.
Our study takes the second approach, drawing on a large corpus of software mentions~\cite{istrate2022large}.

\subsection{Adoption and persistence of software}

Understanding which tools disciplines adopt, and why those choices persist, is essential context for interpreting how the space of software in science takes shape and evolves.
Many factors drive the adoption of software in science. 
The scholarly literature on the adoption of ideas in science and technologies in other sectors has identified a diverse array of relevant factors. 
The intrinsic features of an innovation — including its merit, usability~\cite{ramiro_algorithms_2018}, type~\cite{denrell_ecology_2020}, association with other innovations~\cite{hallett_public_2019,kenter_ad_2015}, novelty~\cite{chen_representing_2017}, and disruptiveness~\cite{ho_disruptive_2022} — can all affect the speed and depth of adoption. 
External effects such as endorsements by prominent early adopters~\cite{merton_matthew_1968,doehne_how_2023,morgan_prestige_2018}, social embeddedness~\cite{berwick_disseminating_2003,joppa_troubling_2013,uzzi_collaboration_2005,latour_science_1987}, and the structure of the underlying social network~\cite{granovetter_threshold_1978,valente_network_2012} also play a vital role. 
Physical tools face additional adoptive constraints, requiring costly equipment~\cite{free_achieving_2004} or specialized training~\cite{gemunden_role_2007}; software is somewhat different, as scientists typically acquire software skills from peers and self-study rather than formal training~\cite{hannay2009scientists}, meaning tool choices propagate primarily through local communities. 
What is clear from this literature is that while merit is a necessary component of the decision to adopt, it is by no means sufficient.

Just as adoption of software tools is shaped by social forces, so too is their persistence. 
The genomics community's continued reliance on \texttt{Excel} persists despite evidence that its default settings erroneously convert gene names like SEPT2 into dates, an problem that persists years after it was widely publicized~\cite{ziemann2016gene, abeysooriya2021gene}. 
In immunology, flow cytometry analysis depends on software like \texttt{FlowJo} that is tightly coupled to cytometer hardware and proprietary file formats, constraining tool choice through infrastructure rather than preference~\cite{white2021flowkit}. 
In machine learning, \texttt{Python} has achieved overwhelming dominance, a position now reinforced as large language models (LLMs) themselves preferentially generate Python code~\cite{twist2025llms}. 
Software also standardizes practice in ways that extend beyond any single tool: when thousands of researchers run differential expression analysis through \texttt{DESeq2}~\cite{love2014moderated}, they share not just a method but a specific implementation, making findings more directly comparable --- and making departure from that implementation increasingly costly. 
In each case, persistence is explained not by intrinsic quality but by how a tool is embedded in a broader social and technical context. 
Once established in a scientific community, software tends to resist displacement.

Software is also not used in isolation, but combined with other tools into scientific workflows. 
Open-source software has long been characterized by an ``ecosystem'' metaphor~\cite{schueller2022evolving,schueller2024modeling,valiev2018ecosystem}, which captures both the heterogeneity of software and its interconnectedness through explicit dependencies and co-usage. 
In scholarly practice, it is common for multiple tools to be combined within a single analysis~\cite{li2018comention}: a biologist might use \texttt{Snakemake} to orchestrate a pipeline in which \texttt{STAR} aligns sequencing reads, \texttt{DESeq2} performs differential expression analysis, and \texttt{ggplot2} produces the final figures --- each tool chosen not just on its own merits but for its compatibility with the others. 
Just as software adoption is best understood at the level of communities rather than individuals, the output of that adoption is best understood not as a collection of individual tool choices but as a \textit{tool portfolio}: a structured set of tools that reflects a discipline's collective workflows, skills, and constraints, and that, like the adoption patterns that produce it, resists change from the outside.

To characterize disciplinary software portfolios, we need a framework that captures portfolios of related tools, rather than the adoption of isolated tools by individual scholars.
Our framework draws on a tradition in complexity economics that maps capability relatedness through co-occurrence patterns in observed outcomes~\cite{hidalgo2007product}.
This approach has been extended to science: Miao et al.~\cite{miao2022latent} applied the same logic to national research portfolios, treating scientific disciplines as ``products" of nations and showing that disciplinary relatedness structures scientific development in ways that parallel economic diversification. Related work has mapped the ``research space" to predict how individuals and institutions develop new research areas~\cite{guevara2016research}. 
We adapt this logic to construct a ``software space'' in which tools are connected based on their co-adoption across scientific disciplines.
We use this framework because simple co-occurrence is dominated by the most widely used tools by virtue of their frequency, which obscures patterns among more niche, but still important, tools.
``Proximity'', a measure of how consistently two tools are co-adopted across disciplines, addresses this by comparing specialization profiles, surfacing meaningful associations that raw counts would obscure.
This map allows us to characterize each discipline not by any single tool but by its position in a shared software space that captures its characteristic tool portfolio.

%
%
\section{Methods}\label{sec3}
\subsection{Data Sources}\label{subsec3.1}
\paragraph{CZ Software Mentions Dataset} 
We sourced data on software mentions from the CZ Software Mentions Dataset, a large-scale corpus of software mentions extracted from 20 million scientific articles~\cite{istrate2022large}. 
The dataset was constructed through a multi-stage pipeline. 
Software mentions were first extracted from full-text articles using a named entity recognition model based on SciBERT~\cite{beltagy2019scibert}, and fine-tuned on the SoftCite dataset~\cite{du2021softcite}. 
The extraction covered two major corpora: 3.8 million papers from the NIH PubMed Central Open Access collection and 16 million papers provided by publishers. 
Raw mentions were then disambiguated using a clustering algorithm, mapping 1.12 million unique mention strings to 97,600 distinct software entities. 

This dataset offers several advantages for our analysis. 
Unlike citation-based approaches that capture only formal software citations, the CZ Software Mentions Dataset identifies informal mentions in methods and analysis sections, providing a more complete picture of actual tool usage~\cite{howison2016software}. 
The disambiguation process also addresses a key challenge in software mention analysis, as the same tool may be referenced through multiple name variants (e.g., ``Scikit-learn'' and ``sklearn''). 
This dataset has been validated and used in subsequent research on software citation practices~\cite{druskat2024don} and software citation intent classification~\cite{istrate2024scientific}.

\paragraph{Dimensions} 
We enriched the CZ dataset with metadata from Dimensions, a bibliographic database integrating publications, citations, grants, and other research outputs~\cite{hook2018dimensions}. 
We use Dimensions' implementation of the Australian and New Zealand Standard Research Classification (ANZSRC) Fields of Research (FoR) codes to assign disciplinary affiliations to papers. 
The FoR classification system comprises a three-level hierarchy: Divisions (2-digit codes representing broad subject areas such as ``Biological Sciences'' or ``Information and Computing Sciences''), Groups (4-digit codes representing more specific fields such as ``Genetics'' or ``Data Management and Data Science''), and Fields (6-digit codes for highly specific topics)~\cite{porter2023recategorising}. 
In our analysis, we used 2-digit Division codes for broader comparisons and 4-digit Group codes for finer granularity in distinguishing disciplinary tool portfolios.
To facilitate analysis, we also defined a higher-level disciplinary category, grouping each of the 22 ANZSRC divisions into one of the three broad categories: Natural and Health Sciences (6 divisions), Physical and Technical Sciences (5 divisions), and Social Sciences and Humanities (11 divisions) (see Table~\ref{tab:categories} in Appendix~\ref{app:categories}). 

Before linking datasets, we performed several preprocessing steps on the CZ Software Mentions Dataset. 
The dataset assigns each mention one of four labels based on their level of curation: \texttt{software}, \texttt{not\_software}, \texttt{unclear}, and \texttt{not\_curated}. 
We retained all mentions labeled \texttt{software} and additionally recovered \texttt{not\_curated} entries whose in-publication text exactly matched known software names from the curated set, increasing coverage by approximately 261,000 mention records. 
We then disambiguated software names in two stages. 
First, we resolved case variants by grouping names that differ only in capitalization (e.g., \texttt{Flowjo} and \texttt{FlowJo}) and mapping all variants to the most frequently occurring form. 
Second, we applied a manually curated mapping table to merge remaining synonyms and aliases into canonical names.
After disambiguation, the dataset contains 5,208 unique software tools mentioned across 1.34 million unique publications.

We matched records between the preprocessed CZ dataset and Dimensions using Digital Object Identifiers (DOIs). 
Papers without DOIs or those not indexed in Dimensions were excluded. 
We further restricted the analysis to publications from 2004 to 2021 to ensure consistent temporal coverage. 
After deduplicating at the paper--software level (counting each software at most once per paper), we retained software whose total mention count exceeded the 90th percentile of the corpus-wide distribution (approximately 990 mentions) in order to focus analysis of software with adequate representation for analysis. 
This yielded 520 tools representing 22 field divisions.

\subsection{Revealed comparative advantage}\label{subsubsec3.2.1}

To identify which software tools are disproportionately associated with which disciplines, we adapt the revealed comparative advantage (RCA) index~\cite{balassa1965trade, laursen2015revealed}. Originally developed to measure countries' export specializations, RCA has been applied in bibliometric research to characterize national research strengths~\cite{miao2022latent}. 
We calculate RCA for all combinations of disciplines and software as:

\begin{equation}
RCA_{d,s} = \frac{\mathcal{M}(d,s) / \sum_{s} \mathcal{M}(d,s)}{\sum_{d} \mathcal{M}(d,s) / \sum_{d,s} \mathcal{M}(d,s)}
\end{equation}

where $\mathcal{M}(d,s)$ denotes the number of papers in discipline $d$ mentioning software $s$, $\sum_{s} \mathcal{M}(d,s)$ is the total number of software mentions in discipline $d$, $\sum_{d} \mathcal{M}(d,s)$ is the total mentions of software $s$ across all disciplines, and $\sum_{d,s} \mathcal{M}(d,s)$ is the grand total of all mentions.

The RCA index compares a discipline's intensity of using a particular tool against the overall intensity of that software's usage across all disciplines. Values greater than 1 indicate that discipline $d$ uses software $s$ more frequently than the scientific average, suggesting a specialization or ``advantage'' in that tool.

\subsection{Software proximity network}
Building on the product space methodology developed by Hidalgo et al.~\cite{hidalgo2007product}, we construct a proximity network in which software tools are nodes, connected based on the degree to which they are co-adopted as specializations across disciplines.
The underlying intuition is that tools requiring similar methodological requirements or serving complementary functions are likely to co-occur as specializations within the same discipline.

We define the proximity $\Phi_{i,j}$ between two software tools $i$ and $j$ as the minimum of their conditional probabilities of co-specialization:
\begin{equation}
\Phi_{i,j} = \min \{P(\text{RCA}_{i} > 1 \mid \text{RCA}_{j} > 1), \; P(\text{RCA}_{j} > 1 \mid \text{RCA}_{i} > 1)\}
\end{equation}

This definition follows the original product space formulation, which takes the minimum conditional probability to ensure symmetry and produce a conservative estimate of relatedness. 
Two tools are considered proximate if disciplines that specialize in one tool tend to also specialize in the other. 
We compute both RCA and proximity using mentions aggregated across all papers in the study period (2004--2021).
The resulting network captures the space of scientific software, where tools cluster based on shared disciplinary usage patterns rather than predefined software categories. 

To identify groups of functionally related tools, we apply a nested stochastic block model~\cite{peixoto2014hierarchical} to the full proximity network. Unlike modularity-based methods such as Louvain or Leiden, the stochastic block model does not require a resolution parameter and can detect communities at multiple scales. 
We use the implementation in \texttt{graph-tool}~\cite{peixoto_graph-tool_2014}. 
This model identifies eight communities at the top level of the hierarchy, which we label based on manual inspection of their constituent tools.

The full proximity network is dense. To produce a readable layout, we extract the network's statistically significant edges using the disparity filter~\cite{serrano2009extracting} and overlay a maximum spanning tree~\cite{kruskal1956shortest} to ensure global connectivity. The resulting backbone retains the most meaningful connections while preserving the network's overall structure.

\subsection{Concentration and stability}
To measure the level of concentration in a disciplinary software portfolio, we draw on the Herfindahl-Hirschman Index (HHI)~\cite{rhoades1993herfindahl}, a standard measure of market concentration, which we use here to assess whether a discipline's software portfolio is concentrated in few communities or spread across many.

\begin{equation}
HHI_{d,t} = \sum_{c \in \mathcal{C}} s_{c,d,t}^2 \quad \text{where} \quad s_{c,d,t} = \frac{n_{c,d,t}}{\sum_{c' \in \mathcal{C}} n_{c',d,t}}
\end{equation}

where $\mathcal{C}$ is the set of detected communities in the proximity network, $n_{c,d,t}$ is the number of software tools in community $c$ used by discipline $d$ during time period $t$, and $s_{c,d,t}$ is the share of community $c$ within discipline $d$'s software portfolio. 
High HHI values (approaching 1.0) indicate concentration in few tool communities, while low values indicate a diverse portfolio spanning multiple communities. 
We compute HHI using a rolling 5-year window to smooth annual fluctuations and restrict this calculation to tools with RCA $> 1$ in the given division and time window, so that HHI reflects the concentration of a division's specialized tools rather than all tools mentioned.

To quantify how stable each discipline's software portfolio is over time, we measure the Jaccard similarity~\cite{niwattanakul2013using} between the specialized tool sets of consecutive rolling 5-year windows:

\begin{equation}
\mathcal{J}_{d,t} = \frac{|S_{d,t} \cap S_{d,t+1}|}{|S_{d,t} \cup S_{d,t+1}|}
\end{equation}

where $S_{d,t}$ is the set of software tools with RCA $> 1$ for discipline $d$ in year $t$, and $S_{d,t+1}$ is the corresponding set in the next window. 
Values near 1 indicate high portfolio stability, with a division retaining most of its specialized tools from one window to the next; values near 0 indicate rapid turnover.
We report the mean Jaccard similarity across divisions within each broad disciplinary category.

%
%
\section{Results}\label{sec4}
\subsection{The landscape of scientific software}\label{subsec4.1}

The distribution of software mentions follows a power law (see Appendix~\ref{fig:power_law}), indicating that scientific software usage is dominated by a small number of widely adopted tools while the vast majority appear rarely.
The 20 most frequently mentioned tools account for approximately 35\% of all mentions (4 million), led by \texttt{SPSS}, \texttt{R}, \texttt{GraphPad Prism}, \texttt{ImageJ}, and \texttt{Excel} (Table~\ref{tab:top_software}).
The long tail of software beyond these top 20 is predominantly tools used in biomedical workflows. 

\begin{table}[h]
\centering
\caption{20 most frequently mentioned software tools.}
\label{tab:top_software}
\small
\begin{tabular*}{\textwidth}{@{\extracolsep{\fill}}lr lr@{}}
\toprule
Software & Mentions & Software & Mentions \\
\midrule
\texttt{SPSS}           & 329,080 & \texttt{ClustalW}       & 23,212 \\
\texttt{R}              & 186,737 & \texttt{Adobe Photoshop}& 22,724 \\
\texttt{GraphPad Prism} & 181,411 & \texttt{SHELXL97}       & 20,754 \\
\texttt{ImageJ}         & 122,008 & \texttt{Statistica}     & 18,539 \\
\texttt{Excel}          &  86,292 & \texttt{Ensembl}        & 18,493 \\
\texttt{Stata}          &  77,365 & \texttt{BLASTN}         & 18,323 \\
\texttt{MATLAB}         &  75,174 & \texttt{Cytoscape}      & 17,839 \\
\texttt{BLAST}          &  71,582 & \texttt{SigmaPlot}      & 17,810 \\
\texttt{MEGA}           &  36,276 & \texttt{SHELXS97}       & 17,557 \\
\texttt{FlowJo}         &  31,577 & \texttt{Bowtie}         & 17,492 \\
\bottomrule
\end{tabular*}
\vspace{0.5em}
\parbox{\textwidth}{\footnotesize\textit{Note.} \texttt{BLAST} and \texttt{BLASTN} are listed separately. \texttt{BLASTN} is a nucleotide-specific module within the \texttt{BLAST} suite. We retain distinct entries for suite components throughout the analysis. Similarly, \texttt{SHELXL97} and \texttt{SHELXS97} are separate programs within the \texttt{SHELX} suite, used for structure refinement and structure solution, respectively.}
\end{table}

These aggregate counts, however, tell us little about how usage varies across disciplines. 
Some tools are genuinely ubiquitous: \texttt{R} appears in 22 divisions (167 of 170 4-digit groups) with no single field accounting for more than 12\% of its usage. 
Others serve far narrower communities: \texttt{BEAST} concentrates 66\% of its usage in just three fields (Genetics, Evolutionary Biology, and Ecology), while \texttt{GROMACS} draws 58\% of its mentions from Biochemistry and Cell Biology, Medicinal Chemistry, and Bioinformatics. 
A simple ranking of popular tools thus provides limited insight into how software is incorporated into scholarly practice.

\begin{figure}[h]
\centering
\includegraphics[width=0.95\textwidth]{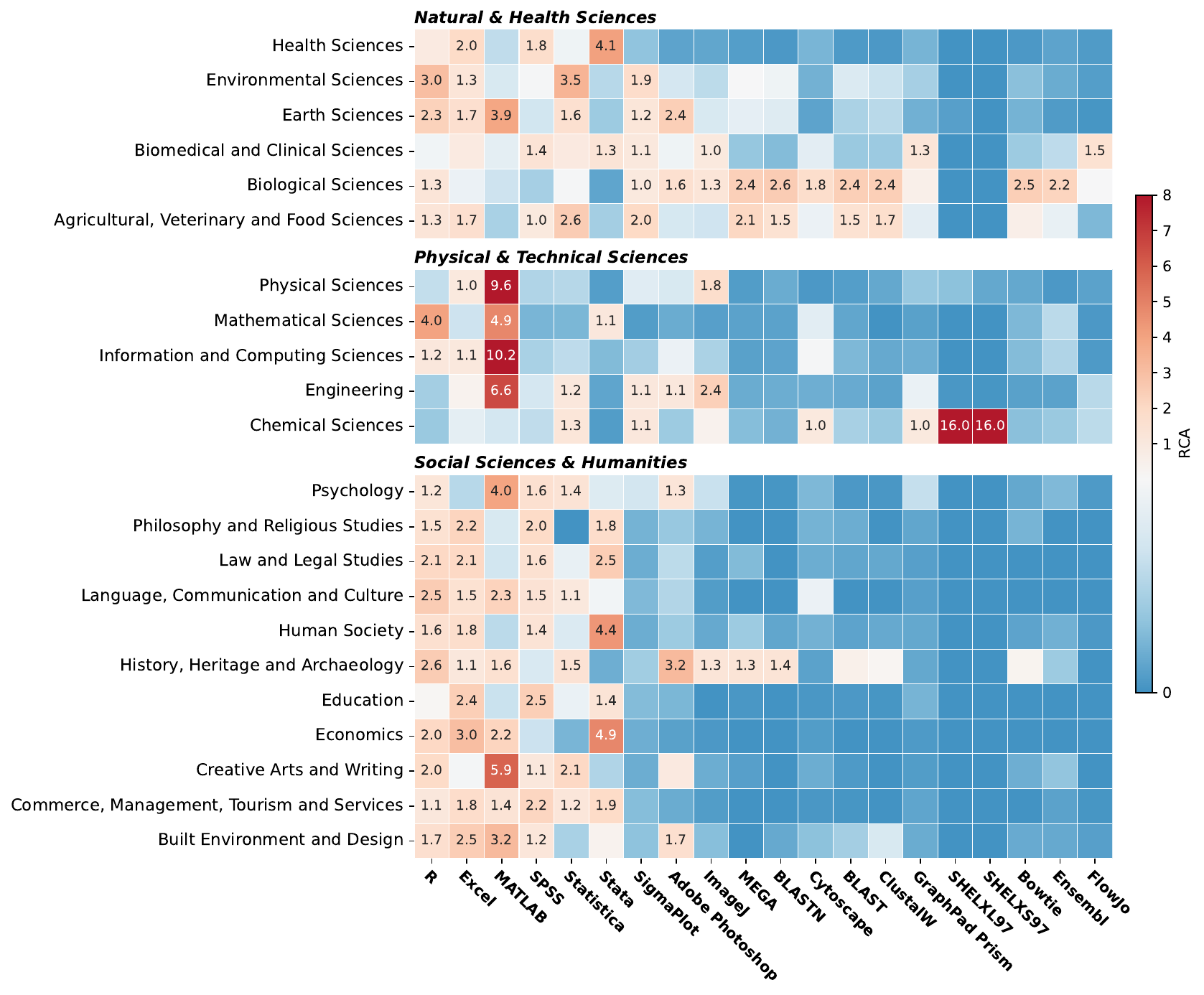}
\caption{
\textbf{How disciplines specialize in software.} 
Shown is the revealed comparative advantage (RCA) for the 20 most popular software and each of the 22 disciplinary divisions. 
Divisions are grouped into three broad categories (Natural \& Health Sciences, Physical \& Technical Sciences, Social Sciences \& Humanities). 
Red cells (RCA $\geq 1$) indicate above-average mention intensity or a ``specialization''; blue cells (RCA $< 1$) indicate below-average usage. 
Software is ordered left to right by decreasing number of divisions with RCA $\geq 1$.}\label{fig:rca_heatmap_top_software}
\end{figure}

To reveal how the same tools take on different roles across disciplines, we compute the RCA of each tool within each broad disciplinary division. Fig.~\ref{fig:rca_heatmap_top_software} displays RCA values for the 20 most mentioned tools across 22 divisions. Software is ordered from left to right by the number of divisions in which it achieves RCA $\geq 1$, so that generalist tools appear on the left and those more (relatively) specialist tools on the right. 

The left side of the heatmap is dominated by statistical and computational platforms. \texttt{R} has by far the most disciplines in which it is classed as specialized, spanning most field divisions with few exceptions such as Biomedical and Clinical Sciences and Engineering. 
Next are \texttt{Excel}, \texttt{MATLAB}, and \texttt{SPSS}. 
\texttt{MATLAB} and \texttt{SPSS} achieve RCA $\geq 1$ in the same number of divisions, but their profiles barely overlap: \texttt{MATLAB} peaks in the Physical Sciences and Information \& Computing Sciences, while \texttt{SPSS} concentrates in the Social Sciences and Humanities. 
The right side, representing those with fewer disciplinary divisions marked as specialized, quickly becomes sparse. 
\texttt{SHELXL97} and \texttt{SHELXS97} reach extreme RCA values in Chemical Sciences but do not appear in any other division. Bioinformatics tools such as \texttt{BLAST}, \texttt{ClustalW}, \texttt{Bowtie} and \texttt{Ensembl} show RCA above 1 only in the biological and agricultural sciences. 

To move beyond the most popular tools and map the broader landscape, we next examine how all 520 tools relate to one another through disciplinary co-specialization.

%
\subsection{Tool portfolios}\label{subsec4.2}
\begin{figure}[h]
\centering
\includegraphics[width=\textwidth]{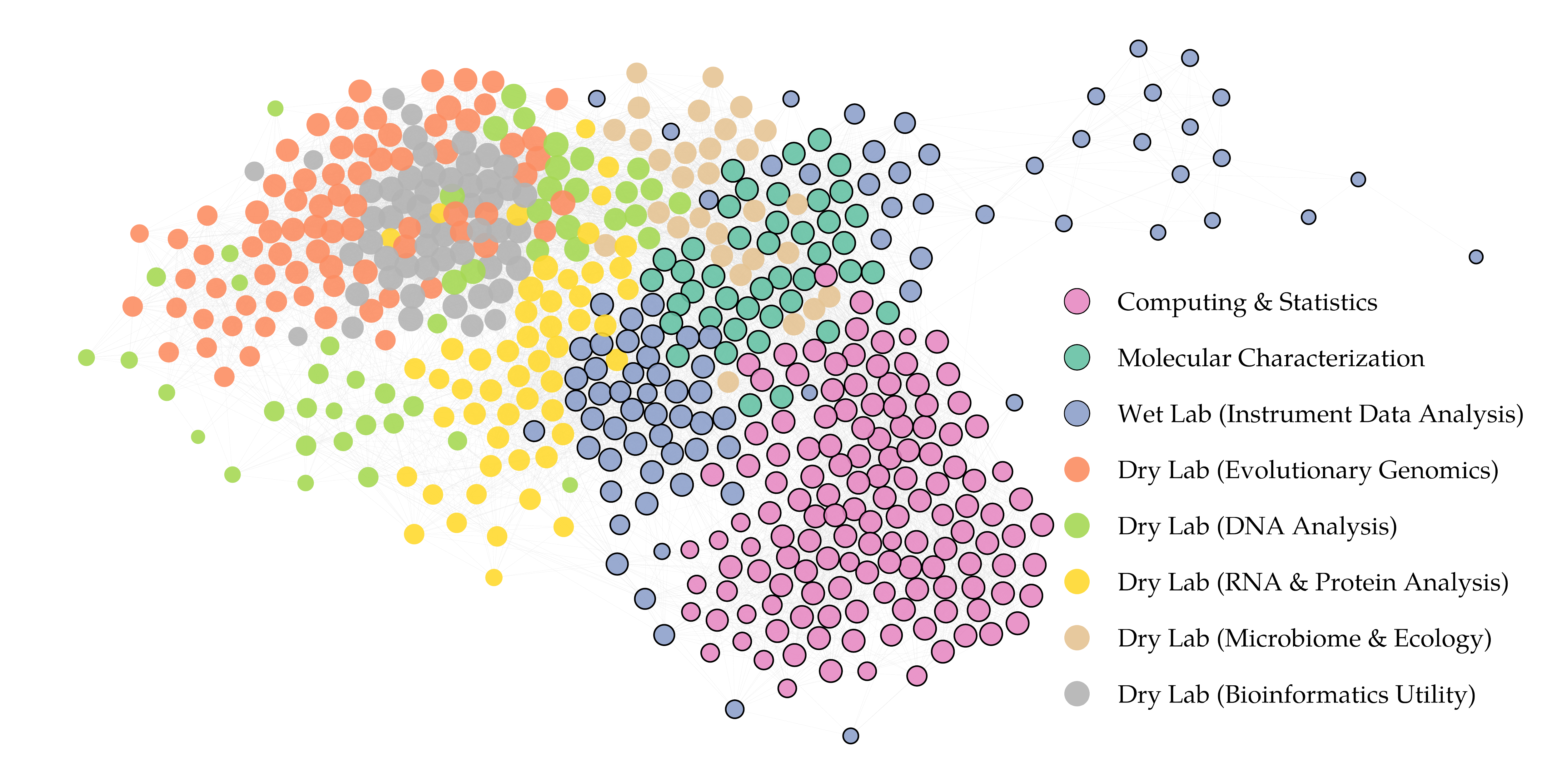}
\caption{
\textbf{The software space of science reveals a structured landscape of tools.}
Shown is a network ($V = 520$, $E = 10{,}180$) of software tools connected by disciplinary co-specialization ($RCA > 1$). Node colors indicate eight communities identified by a nested stochastic block model. The layout displays the network's weighted backbone~\cite{serrano2009extracting} with a maximum spanning tree~\cite{kruskal1956shortest} overlaid to preserve global connectivity. Wet lab tools, tied to laboratory instrumentation, anchor one end of the network, while dry lab tools fragment into specialized subcommunities for evolutionary genomics, DNA, RNA \& protein, microbiome \& ecology, and bioinformatics utility. The Molecular Characterization community bridges experimental and computational work. The Computing \& Statistics community forms a distinct cluster, reflecting general-purpose tools shared across domains.}\label{fig:software_space}
\end{figure}

Rather than reflecting independent adoption, software tools cluster into structured communities when linked by disciplinary co-specialization. 
Using RCA to identify specializations and proximity to measure co-specialization patterns, we construct a network linking tools that share comparative advantage across disciplines. 
The resulting structure (Fig.~\ref{fig:software_space}) reveals a software space structured by patterns of shared disciplinary use.

The resulting network does not exhibit the core-periphery structure characteristic of the ``product space'' of national exports~\cite{hidalgo2007product}, nor the small number of well-defined communities observed in the ``research space'' of national publications~\cite{miao2022latent}, instead exhibiting a domain-specific structure with loose boundaries. 
The most striking division separates wet lab from dry lab tools. 
The \textit{Wet Lab} community (85 tools) anchors one end of the network, containing software tied to specific laboratory instruments such as \texttt{FlowJo} for flow cytometry, \texttt{SHELXL97} for crystallography, and \texttt{ZEN} for microscopy, as well as widely used biomedical analysis tools like \texttt{GraphPad Prism}. 
Adjacent to it, the \textit{Molecular Characterization} community (43 tools) sits at the boundary between experimental and computational work. 
It brings together mass spectrometry software (\texttt{Mascot}, \texttt{Xcalibur}), qPCR analysis tools (\texttt{geNorm}, \texttt{CFX Manager}), and molecular simulation tools (\texttt{GROMACS}, \texttt{AMBER}, \texttt{AutoDock}), reflecting workflows where experimental measurements feed directly into computational modeling. 
Unlike the high-throughput omics communities, these tools typically target specific molecules or gene candidates rather than genome-wide discovery.

At the opposite end, dry lab communities split into specialized domains. 
\textit{Evolutionary Genomics} (76 tools) covers sequence alignment and tree building (\texttt{MEGA}, \texttt{RAxML}, \texttt{MrBayes}). 
\textit{DNA Analysis} (48 tools) is oriented around variant calling and population genomics (\texttt{PLINK},  \texttt{GATK}). 
\textit{RNA \& Protein} (52 tools) captures the pipeline from differential expression through pathway analysis and network analysis (\texttt{DESeq2}, \texttt{limma}, \texttt{Cytoscape}). 
\textit{Bioinformatics Utility} (53 tools) provides shared infrastructure across these communities, with tools for search, alignment, and quality control (\texttt{BLAST}, \texttt{SAMtools}). 
\textit{Microbiome \& Ecology} (30 tools) combines amplicon sequencing pipelines (\texttt{QIIME}, \texttt{Mothur}) with ecological statistics software (\texttt{vegan}, \texttt{PAST}).

The \textit{Computing \& Statistics} community (133 tools) is the largest and spans from the wet lab to the dry lab side of the network. 
It includes general statistical software (\texttt{SPSS}, \texttt{Stata}), programming languages (\texttt{R}, \texttt{Python}, \texttt{MATLAB}), and image analysis tools (\texttt{ImageJ}, \texttt{Fiji}). 
That this community bridges the two sides of the network rather than clustering with either suggests that shared computational infrastructure 
connects disciplines across, as well as within, subject domains.

\begin{figure}[!htbp]
\centering
\includegraphics[width=0.85\textwidth]{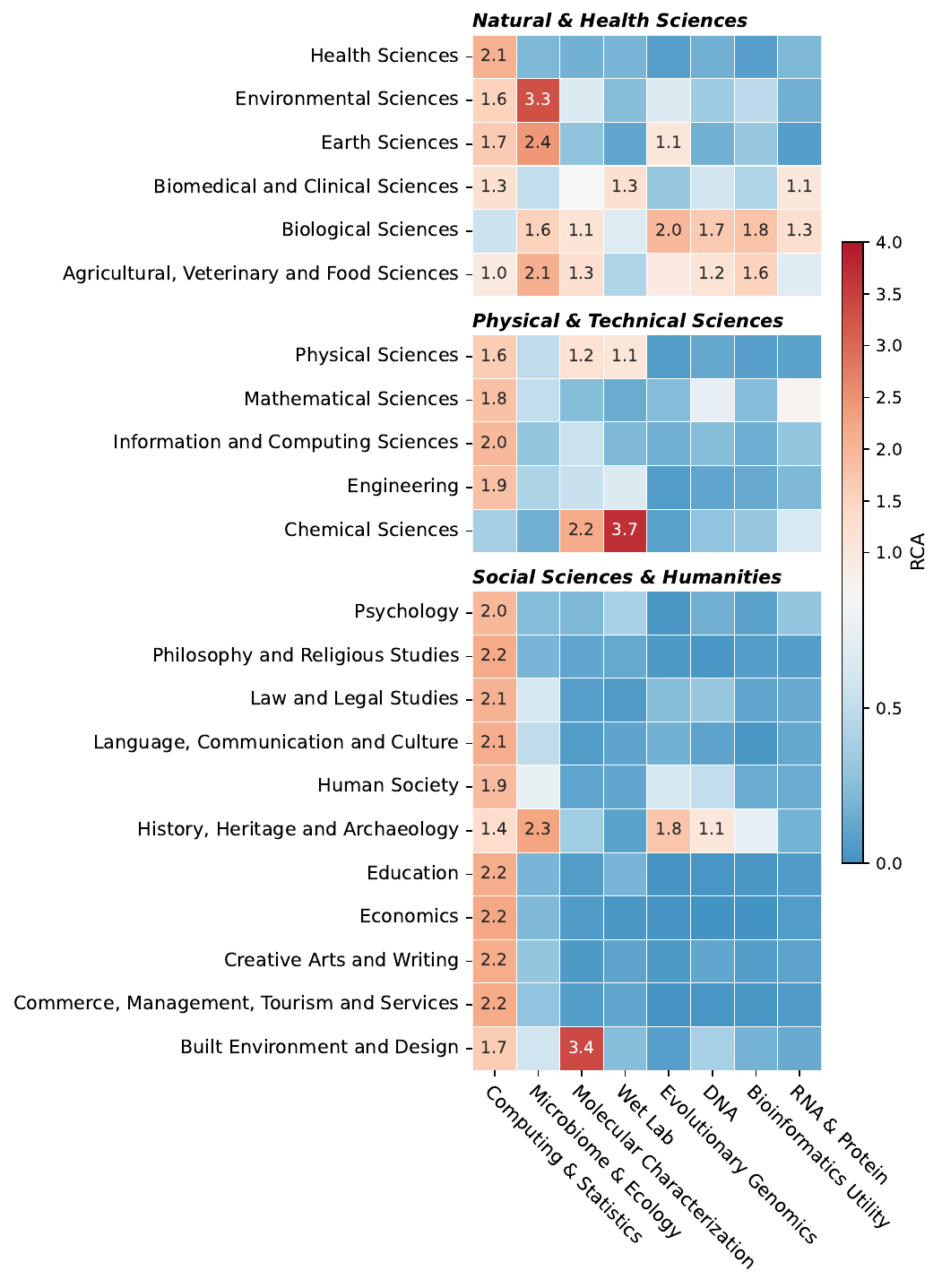}
\caption{
\textbf{Disciplinary tool portfolios.}
RCA of 22 disciplinary divisions across software communities computed over the full study period (2004--2021).
For each division, we sum the mentions of all tools belonging to a given community, producing a division $\times$ community matrix, and then compute RCA on this matrix.
Divisions are grouped into three broad categories (Natural \& Health Sciences, Physical \& Technical Sciences, Social Sciences \& Humanities). Red cells (RCA $\geq 1$) indicate above-average specialization in a community; blue cells (RCA $< 1$) indicate below-average specialization.}\label{fig:rca_heatmap_community}
\end{figure}

We compute the RCA of each disciplinary division against each of the 8 tool communities (Fig.~\ref{fig:rca_heatmap_community}) to show how disciplinary tool portfolios are positioned in this software space. 
Within Natural and Health Sciences, disciplines differ in the breadth of their portfolios. 
Biological Sciences achieves RCA above 1 in six of eight communities, spanning all biology-oriented communities, but falls below threshold in both Wet Lab and Computing \& Statistics. 
This suggests that compared to other divisions, biological research is distinguished more by its specialized analytical tools than by general computational infrastructure. 
Environmental Sciences and Earth Sciences both concentrate heavily in Microbiome \& Ecology, with Earth Sciences also drawing on evolutionary genomics tools.
Agricultural Sciences shows a similarly broad profile to Biological Sciences. 
By contrast, Health Sciences and Biomedical and Clinical Sciences both rely primarily on Computing \& Statistics, with Biomedical Sciences additionally drawing on Wet Lab tools and RNA \& Protein analysis.

Among Physical and Technical Sciences, Chemical Sciences stands out for its extreme concentration in just two communities: Wet Lab and Molecular Characterization, with everything else well below threshold. 
This is a field built around instruments and molecular-level analysis, largely disconnected from the sequence-oriented and ecological tools that dominate the life sciences. 
Physical Sciences is specialized in three communities, combining computing tools with wet lab and molecular characterization. 
Engineering, Mathematical Sciences, and Information and Computing Sciences each concentrate almost entirely in Computing \& Statistics.

Most Social Sciences and Humanities divisions rely almost entirely on general purpose computing and statistical tools. 
Psychology, Philosophy, Law, Education, Economics, and several others all exceed an RCA of 2.0 in Computing \& Statistics while falling well below 1 everywhere else. 
The notable exception is History, Heritage and Archaeology, which shows specialization in 4 communities including Microbiome \& Ecology and Evolutionary Genomics. 
Inspection of the underlying tools reveals this is driven by paleontological statistics software and ancient DNA analysis tools, reflecting the growing integration of molecular and ecological methods into archaeological research. 
Built Environment and Design shows an unexpectedly high RCA in Molecular Characterization, but this appears to be a classification artifact: the underlying papers are computational drug design studies misclassified under the ``Design" subfield.

%
\subsection{Structural differences and stabilization of disciplinary tool portfolios}\label{subsec4.3}

To examine how tool portfolios vary across disciplines and over time, we compute two measures for each division over rolling 5-year windows from 2004 to 2021. 
The HHI captures the concentration of a division's software usage across tool communities. Values near 1 indicate reliance on a single community, while lower values indicate usage spread across several. 
The Jaccard similarity between consecutive windows measures how much of a division's specialized software portfolio (the set of tools with RCA $> 1$) carries over from one period to the next.

\begin{figure}[h]
\centering
\includegraphics[width=0.9\textwidth]{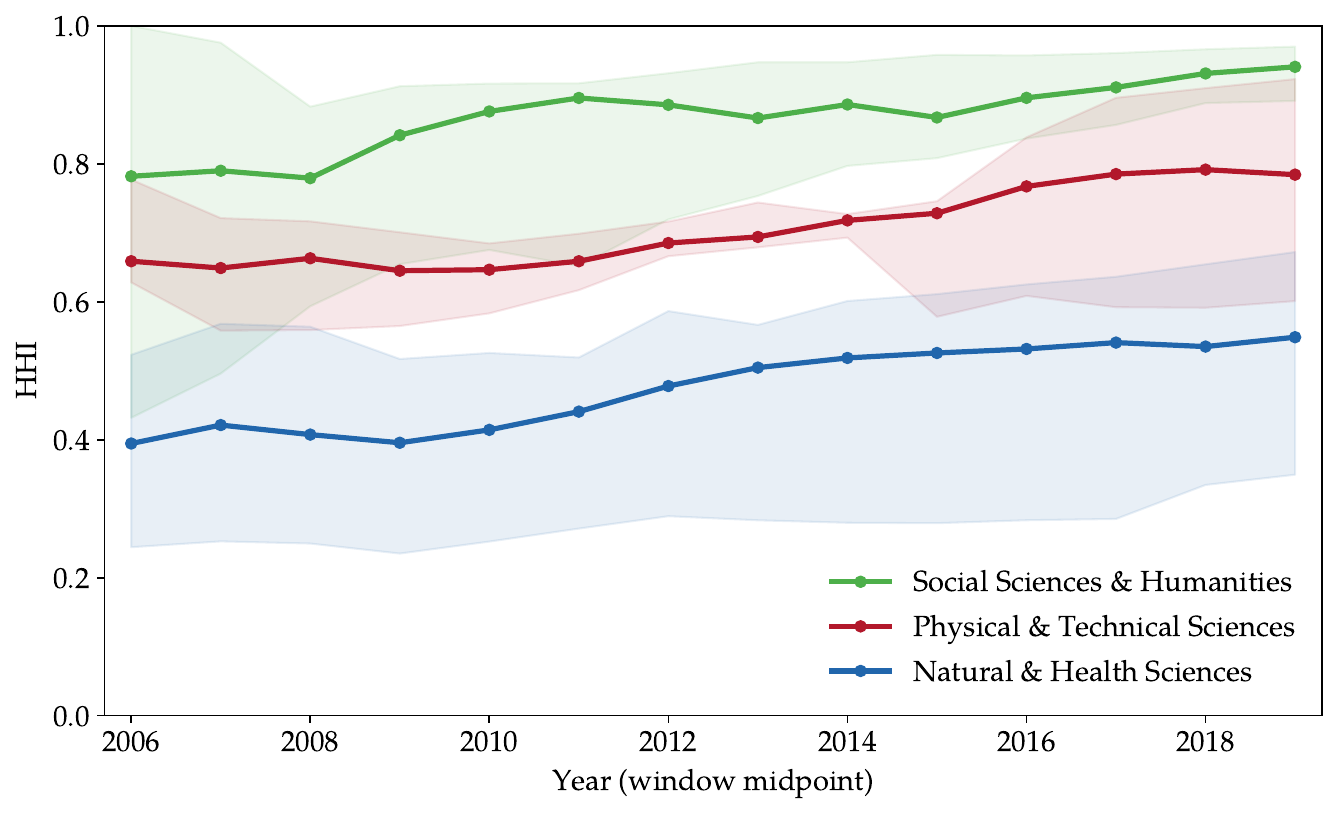}
\caption{
\textbf{Software community concentration rises across all three disciplinary categories, but disciplinary gaps remain.}
Median HHI across divisions within each category, computed over rolling 5-year windows from 2004--2008 to 2017--2021. 
Shaded bands show the interquartile range. 
The underlying publication volumes over the full study period differ substantially across categories (Natural \& Health Sciences: 1.05M papers; Physical \& Technical Sciences: 132K; Social Sciences \& Humanities: 64K), and HHI estimates for smaller categories may be more volatile due to lower publication counts.}\label{fig:hhi_trajectory}
\end{figure}

Fig.~\ref{fig:hhi_trajectory} plots the median HHI within three broad disciplinary categories over time.
The most prominent feature is the persistent gap between categories. 
Natural and health sciences maintain the lowest concentration throughout (median HHI rising from 0.39 to 0.55), consistent with workflows that draw on multiple tool communities, from wet lab and bioinformatics to computing and statistics. 
Social sciences and humanities remain the most concentrated (0.78 to 0.94), with software usage dominated by a single community: computing and statistical analysis. Physical and technical sciences fall in between (0.66 to 0.78). 
Although all three categories shift upward over time, the ranking and relative spacing between them remain stable, suggesting that a discipline's tool breadth is shaped more by its research workflow than by any temporal trend.

\begin{figure}[h]
\centering
\includegraphics[width=0.95\textwidth]{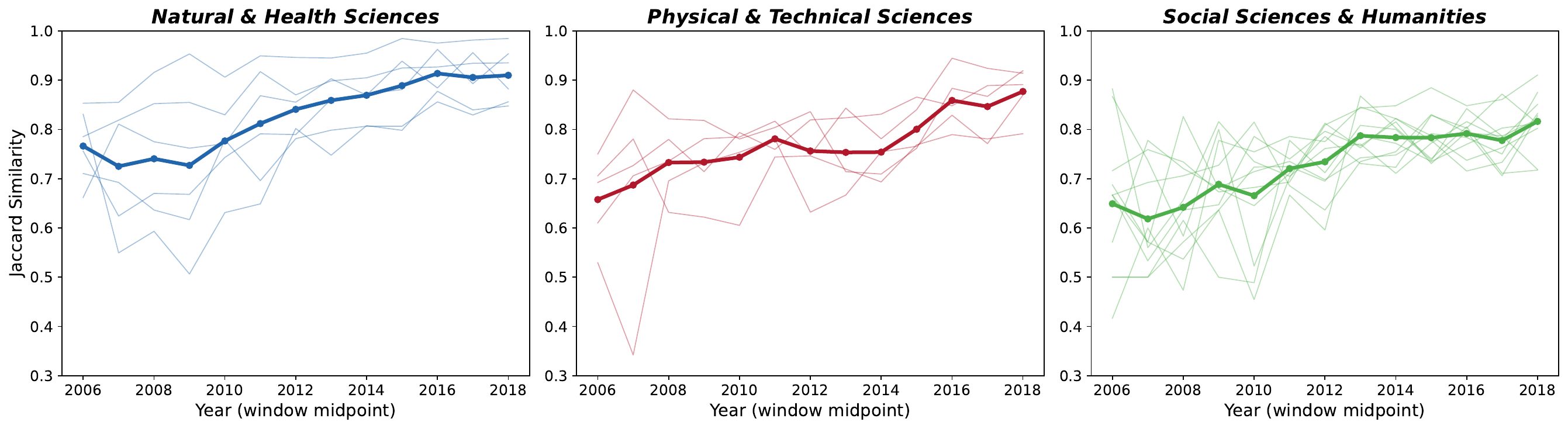}
\caption{
\textbf{Software portfolios are stabilizing over time.}
Stability of disciplinary software portfolios, measured as the Jaccard similarity between the sets of specialized tools (RCA $> 1$) in consecutive rolling 5-year windows. Thin lines show individual divisions; thick lines show the mean within each broad category.}\label{fig:jaccard}
\end{figure}

Jaccard similarity, which measures the overlap in a division's specialized tools between consecutive windows, rises steadily across all three categories (Fig.~\ref{fig:jaccard}). 
In each case, divisions are retaining a larger share of their specialized tools from one window to the next. 
This pattern holds despite the large differences in tool breadth, as Social Sciences and Humanities are stabilizing at a similar rate as in the Natural and Health Sciences. 
Even disciplines with diverse, multi-community toolkits are increasingly retaining the same software from one period to the next.

Individual divisions follow varied trajectories within these broad patterns (see Appendix Fig.~\ref{fig:hhi_appendix} and Fig.~\ref{fig:jaccard_appendix}). 
The clearest exception is Chemical Sciences, whose HHI dropped from 0.90 to 0.29, though this is partly driven by changes in the subfield composition of the division over time rather than individual researchers broadening their tool use. 
(Inorganic Chemistry, which relies almost exclusively on crystallography software, accounted for 95\% of papers in the earliest window but only 10\% in the latest, while Medicinal and Biomolecular Chemistry, which draws on molecular simulation and docking tools, grew from 2\% to 47\%.) 
Nevertheless, disciplinary software portfolios are overall settling into a more stable configuration of tool communities.

%
%
\section{Discussion}\label{sec5}
Understanding how scientific tools are adopted, spread, and evolve is fundamental to understanding how science itself advances.
Software offers a unique window into this phenomenon, being both pervasive across nearly all scientific disciplines and traceable in the published literature at scale. 
Using a corpus of software mentions across millions of scientific articles, we investigate how software is used in science.
Our study makes three major contributions.

First, we introduce the ``product space'' framework~\cite{hidalgo2007product} to the study of tool use in science. 
Where the original framework maps countries to exported products, we map disciplines to software tools, using RCA to identify specialization and proximity to link tools that co-occur across fields, and ``map'' the software space of science.

By itself, RCA demonstrates that certain kinds of software are ubiquitous across many scientific fields, particularly generalist statistical and programming tools like \texttt{R}, \texttt{Excel}, and \texttt{MATLAB}. 
Most software, even certain popular tools, is far more specialized.
For example, \texttt{FlowJo}, a data analysis software for flow cytometry, has over 30,000 mentions in our dataset but only Biomedical \& Clinical Sciences shows a specialization in the tool.
The long tail of software reflects this pattern: tools oriented toward specific disciplinary use cases, largely reflecting the needs of primarily biomedical fields. 

The true value of the product space framework stems from the resulting proximity network, or ``software space'' of science. 
Previous bibliometric work has mapped the organization of science through citations~\cite{boyack2005mapping} and collaboration networks~\cite{newman2004coauthorship}. 
Software use offers a different view. It captures the computational and material infrastructure through which research is conducted, a dimension that is difficult to recover from citation data alone.
The structure of the resulting software space is not the core-periphery structure of the original product space~\cite{hidalgo2007product}, where some products (e.g., machinery) require capabilities that are broadly transferable.
Neither does the network structure resemble the ``research space''~\cite{miao2022latent}, in which a few well-defined clusters mirror national scientific capacities and priorities.
Rather, we observe 8 communities, each roughly comprising a disciplinary workflow, with less well-defined structure.
This fragmented structure suggests that scientific software is more granular than these other domains, with many narrow and modular workflows and few truly generalist tools. 
Those tools that are generalist tend to be programming languages and applications that support data analysis, such as \texttt{R} and \texttt{MATLAB}, which represent a small fraction of all mentioned software tools. 

The network structure also illustrates several striking divisions. 
One is the distinction between ``Dry Lab'' biomedical workflows, where software supports various computational analyses such as sequence alignment, evolutionary genomics, and omics data processing, and ``Wet Lab'' biomedical workflows, characterized by software that supports analysis of data collected from physical laboratory instruments. 
This distinction is widely recognized in laboratory practice but not previously characterized at scale through software usage data. 
The Molecular Characterization community bridges these two modes, combining instrument-linked software for mass spectrometry, qPCR, and NMR spectroscopy with computational tools for molecular modeling and docking.
One of the largest communities encompasses a broad array of generalist statistical, machine learning, and computing tools. Structurally, this community acts as connective tissue between wet lab and dry lab communities; notably, however, it forms a distinct cluster rather than sitting at the network's core. Several communities — among them evolutionary genomics analysis and bioinformatics utilities — show little proximity to it, suggesting that certain workflows have become relatively self-contained, with their own ecosystems of tools suited to their needs.
Our observations also align with Becher and Trowler's~\cite{becher2001academic} taxonomy of disciplinary cultures, which distinguishes ``hard'' fields with convergent methods and shared standards from ``soft'' fields with more diverse and individualistic approaches. 
In our data, this maps onto a clear difference in software portfolios: ``hard'' disciplines maintain specialized tool ecosystems spanning multiple communities, while ``soft'' disciplines rely almost entirely on statistical software. 

The second major contribution this study makes is positioning the particular software specializations of scientific disciplines. 
Prior work has shown that software mention and citation practices vary across disciplines~\cite{pan2016disciplinary, schindler2022role}, but these studies examined tools individually rather than as part of structured portfolios. 
Our community-level analysis reveals how these differences are structured.
Many disciplines show specialization in only one community---in every case the generalist community. 
Disciplines such as Information \& Computing Sciences, Psychology, and Economics tend toward generalist tools like statistical software rather than adopting more specialized workflows. 
Others are not so confined: Biological Sciences, Agricultural, Veterinary and Food Sciences, and History, Heritage and Archaeology all show cross-cutting specializations across multiple communities. 
Chemical Sciences presents yet another pattern, concentrating almost entirely in Wet Lab and Molecular Characterization tools, with minimal engagement elsewhere.

Together, the structure of the software space and disciplinary tool portfolios suggests that scientific tools are organized not by subject matter alone, but by the underlying workflows through which research is conducted. 
Many fields, particularly in the Social Sciences and Humanities, have workflows satisfied by generalist tools for basic computing and data analysis, and tend toward narrower, more concentrated portfolios. 
Other fields, particularly in the Natural and Health Sciences, comprise many types of workflows, from interfacing with instrument data to modeling to specialized data processing, that cut across disciplinary boundaries in ways that citation and collaboration networks do not reveal, producing less concentrated portfolios.

This finding has implications for how tools are adopted.
Prior work has shown that scientists primarily learn about software through self-study and peers rather than formal training, and that most rely on tools with large user bases~\cite{hannay2009scientists, wilson2014best}.
If tool breadth were primarily driven by availability or awareness, we might expect it to converge over time as software becomes easier to access and share. 
Instead, the gap holds steady. 
This suggests that the number of tool communities a field engages with is constrained less by awareness than by its research practice. 
Fields that combine experimental instrumentation with computational analysis need multiple communities, while fields organized around a single analytical tradition do not.

The third major contribution is the empirical observation of stabilization in disciplinary tool portfolios over time. 
All disciplinary divisions show rising stability at comparable rates despite differences in the structure and breadth of their tool portfolios. 
One reading is that fields have converged on effective tools and built cumulative expertise around them.
Researchers train students on specific software, develop shared workflows, and accumulate tacit knowledge that makes switching costly. Under this reading, stabilization reflects maturity. 
An alternative reading draws on the economics of path dependence~\cite{david1985clio, arthur1989competing}. As Arthur showed for competing technologies, early adoption advantages can compound through positive feedback until an incumbent becomes difficult to displace, not because it is superior, but because the costs of coordination around a replacement are too high. 
Switching costs, training investments and community norms may keep existing tools in place even when better alternatives become available~\cite{farrell2007coordination, klemperer1995competition}. 
Analogous patterns appear in programming language adoption, where the availability of libraries, existing codebases, and prior experience, rather than intrinsic language features, predict which languages developers choose~\cite{meyerovich2013empirical}. 
Our data cannot distinguish between these mechanisms. Doing so would require linking tool dynamics to scientific outcomes. 
For example, whether disciplines with more stable toolkits produce more reproducible findings, or whether adopting new tools leads to methodological breakthroughs.

One open question is the extent to which findings regarding software tools generalize to other types of scientific tools. Software is unique. 
Its adoption costs are typically far lower than those of physical instruments, it is instantly distributable, and it requires no physical infrastructure. 
Adopting physical instruments, in contrast, is often subject to procurement costs, hardware dependencies, and in some cases specialized training. 
This makes the stabilization finding more striking. 
In a domain where switching costs are relatively low and alternatives are readily accessible, scientific tool portfolios are crystallizing. 
Whether analogous patterns hold for physical instrumentation remains an open question, but the software case may represent a lower bound on tool entrenchment in science more broadly.

These findings have practical implications for several audiences. 
Several major funders now support research software as infrastructure, including the Chan Zuckerberg Initiative's Essential Open Source Software for Science program (co-funded with the Wellcome Trust and the Kavli Foundation)\footnote{\url{https://chanzuckerberg.com/rfa/essential-open-source-software-for-science/}}, the Alfred P. Sloan Foundation's programs on open source in science\footnote{\url{https://sloan.org/programs/digital-technology/open-source-in-science}}~\cite{strasser2022ten}, and NSF programs such as CSSI\footnote{\url{https://www.nsf.gov/funding/opportunities/cssi-cyberinfrastructure-sustained-scientific-innovation}} and POSE\footnote{\url{https://www.nsf.gov/funding/opportunities/pose-pathways-enable-open-source-ecosystems}}.
For these funders, the software space identifies which tools function as infrastructure and which disciplinary communities they serve, information that can guide these investments. 
The stabilization trend suggests that established tools are difficult to displace, which matters for decisions about where to invest in new tool development. 
For research computing teams at universities, knowing which tool communities different departments rely on can inform training and support. 
For the open science community, standardized toolkits make findings more comparable across labs, but entrenched tools may also slow the uptake of better methods when they appear.
More broadly, the disciplinary clustering of tool portfolios suggests that the demand for diverse tools arises from the structure of the work itself. 
Simply making tools available may not be enough to broaden adoption across disciplines.

We note several limitations of our study.
First, the CZ Software Mentions Dataset is drawn primarily from biomedical and life science corpora. 
Fields that appear tool-poor in our analysis may simply be underrepresented in this corpus, and the specialization patterns we observe in the Social Sciences and Humanities should be interpreted with this asymmetry in mind. 
Software mentions in papers are also an imperfect proxy for actual use, since norms for reporting software vary across disciplines~\cite{pan2016disciplinary} and citation practices for the same tool can be inconsistent even within a single field~\cite{li2019challenges}.
The CZ dataset relies on text extraction from publications, which may miss software mentioned in supplementary materials or code repositories. 
Second, our analytical choices shape the results. 
We restrict attention to tools above the 90th percentile of corpus-wide mentions, excluding emerging or niche software; the RCA threshold of 1 is binary; and community detection via stochastic block models is one algorithm among several. Division-level trends can also reflect compositional shifts in underlying subfields rather than changes in practice, as we showed for Chemical Sciences.
Third, our study period ends in 2021, before the widespread adoption of large language models and AI-assisted coding tools. Whether these technologies will accelerate stabilization by further standardizing practice, or disrupt it by lowering the barrier to unfamiliar tools, is a question our data cannot address.

Future work could extend this framework in several directions. Linking tool dynamics to scientific outcomes, such as reproducibility and impact, would move the study of scientific tools from mapping to explanation. Constructing dynamic software networks, rather than the static aggregate presented here, would reveal how the topology of the software space evolves as new tools emerge and old ones decline.
One particular avenue to investigate in the future is the impact of large language models on software adoption. 
LLM-based coding assistants have been shown to substantially increase programming productivity~\cite{peng2023impact} and to lower barriers for users without formal programming training~\cite{hou2025comparing}.
On one hand, code-generation via large language models can reduce frictions associated with adopting new software, since learning through interaction with a model or generation of code can mitigate the need for specialized technical training.
This could allow users to discover and experiment with new kinds of software tools that they would not have been exposed to, or able to use, if relying on their disciplinary training alone. 
On the other hand, large language models may narrow the scope of software used, recommending the same types of tools to users no matter their background, likely pushing users away from more specialized workflows and toward more generalist tools. 

Scientific progress depends not only on the questions researchers ask, but on the tools through which they ask them. 
By mapping the software space of science, we show that these tools are not chosen freely or independently, but are structured by disciplinary workflows, stabilized by community norms, and increasingly resistant to change. 
Understanding this hidden infrastructure of science---how it is organized, how it persists, and how it might be disrupted---is essential to understanding how science itself evolves.

\backmatter

\clearpage

\section{Declarations}

\bmhead{Availability of data and materials}
The CZ Software Mentions Dataset is publicly available at \url{https://doi.org/10.5061/dryad.6wwpzgn2c}. 
Disciplinary metadata were obtained from the Dimensions bibliographic database. 
The curated dataset and all derived data products used in this study are available at [Zenodo/Figshare DOI placeholder]. 
Code for reproducing all analyses and figures is available at 
\url{https://github.com/[username-placeholder]/[repo-name-placeholder]}.

\bmhead{Competing interests}
The authors declare that they have no competing interests.

\bmhead{Funding}
This work was funded by the National Institute of General Medical Sciences of the National Institutes of Health (NIH) under Grant No.\ R01GM158813 (PI: Albert-L\'aszl\'o Barab\'asi).

\bmhead{Authors' contributions}
ZW performed the analysis and wrote the initial manuscript. 
DM conceived the project, supervised the research, and revised the manuscript. 
All authors approved the final manuscript.

\bmhead{Acknowledgements}
The authors would like to thank Rodrigo Dorantes-Gilardi and Alina Lungeanu for valuable feedback on earlier versions of this manuscript. 
Zhouming Wu thanks Evangelia Panagakou for her mentorship during the Network Science Institute Research Co-op Program at Northeastern University, where early stages of this work were developed. 
The authors also thank participants at Complex Networks 2025 for helpful comments and suggestions. 






\clearpage
\bibliography{sn-bibliography}

\begin{appendices}

\section{Distribution of software mentions}\label{power_law}

\begin{figure}[H]
\centering
\includegraphics[width=0.85\textwidth]{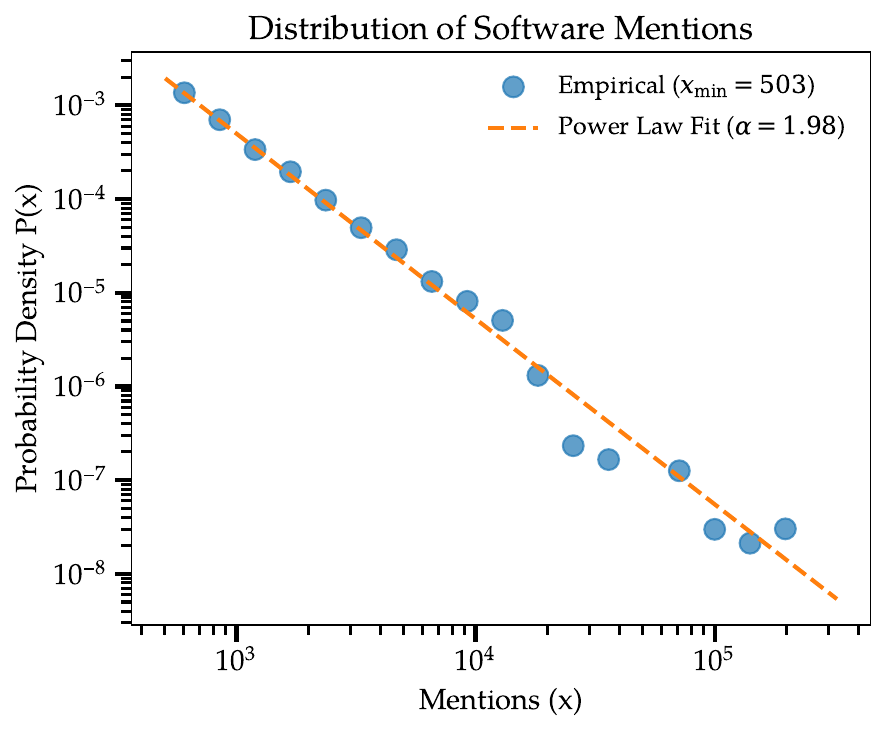}
\caption{Distribution of software mentions across scientific literature (2004--2021). The empirical distribution exhibits a heavy tailed pattern, with a power law exponent $\alpha \approx 1.98$ ($x_{\min} = 503$, the minimum mention count above which the powerlaw fit holds~\cite{clauset2009power}).}\label{fig:power_law}
\end{figure}

\section{Division-to-category mapping}\label{app:categories}

We group the 22 ANZSRC divisions into three broad categories based on the structure of the classification system. Table~\ref{tab:categories} lists the mapping.

\begin{table}[h]
\centering
\caption{Mapping of 22 ANZSRC divisions to three broad categories, with the number of unique papers and software tools in each division.}
\label{tab:categories}
\small
\begin{tabular}{llrr}
\toprule
Category & Division & Papers & Software \\
\midrule
\multirow{6}{*}{Natural \& Health Sciences}
 & Biological Sciences & 356,972 & 528 \\
 & Biomedical and Clinical Sciences & 504,456 & 525 \\
 & Health Sciences & 130,862 & 501 \\
 & Agricultural, Veterinary and Food Sciences & 41,777 & 515 \\
 & Environmental Sciences & 14,146 & 479 \\
 & Earth Sciences & 5,880 & 349 \\
\midrule
\multirow{5}{*}{Physical \& Technical Sciences}
 & Chemical Sciences & 68,493 & 517 \\
 & Engineering & 34,422 & 485 \\
 & Information and Computing Sciences & 22,697 & 479 \\
 & Physical Sciences & 3,510 & 291 \\
 & Mathematical Sciences & 2,832 & 323 \\
\midrule
\multirow{11}{*}{Social Sciences \& Humanities}
 & Psychology & 47,090 & 451 \\
 & Human Society & 5,177 & 339 \\
 & Education & 3,316 & 163 \\
 & Commerce, Management, Tourism and Services & 2,846 & 171 \\
 & Economics & 1,514 & 125 \\
 & History, Heritage and Archaeology & 1,029 & 257 \\
 & Language, Communication and Culture & 925 & 155 \\
 & Built Environment and Design & 730 & 183 \\
 & Law and Legal Studies & 588 & 138 \\
 & Philosophy and Religious Studies & 570 & 96 \\
 & Creative Arts and Writing & 508 & 122 \\
\bottomrule
\end{tabular}
\end{table}

\section{Software communities and their members}\label{app:communities}

The 8 communities identified by the stochastic block model and their member tools, ranked by mention count:

\small
\begin{description}
\item[Molecular Characterization (43)] SigmaPlot, PyMOL, Mascot, Chimera, Primer Express, GROMACS, Xcalibur, Primer-BLAST, MODELLER, geNorm, AMBER, AutoDock, VMD, MassLynx, CHARMM, I-TASSER, CFX Manager, MolProbity, AutoDock Vina, gplots, Discovery Studio, Topspin, SIMCA-P, PROCHECK, NormFinder, LightCycler, GraphPad InStat, MOE, Schr\"odinger, MeV, SIMCA, MassHunter, Rosetta, NAMD, StepOne, ProteinPilot, AutoDockTools, LigPlot, FlexAnalysis, LigPrep, ESPript, LinRegPCR, Empower

\item[Evolutionary Genomics (76)] MEGA, MUSCLE, MAFFT, Perl, Geneious, RAxML, MrBayes, BWA, BioEdit, SPAdes, FigTree, BEAST, Arlequin, Tracer, DnaSP, PAUP*, RepeatMasker, jModelTest, tRNAscan-SE, Velvet, FastTree, RAST, Sequencher, PAML, PHYLIP, Prokka, BWA-MEM, IQ-TREE, Gblocks, TreeAnnotator, GenAlEx, MEGA X, Circos, WebLogo, GENEPOP, Prodigal, SnpEff, SOAPdenovo, Glimmer, OrthoMCL, RNAmmer, AUGUSTUS, Modeltest, Mauve, STRUCTURE HARVESTER, MUMmer, Newbler, RDP, QUAST, CODEML, Mesquite, FSTAT, SeqMan, Lasergene, BioNJ, SplitsTree, antiSMASH, ProtTest, SeaView, trimAl, HHpred, PartitionFinder, adegenet, GeneMark, TargetP, Micro-Checker, Canu, ABySS, Pilon, CLUMPP, MIRA, MrModeltest, RepeatModeler, Unicycler, minimap2, ModelFinder

\item[Wet Lab (85)] GraphPad Prism, FlowJo, SHELXL97, SHELXS97, SAINT, SHELXTL, CellQuest, Quantity One, IPA, MedCalc, PLATON, ORTEP-3, BD FACSDiva, ZEN, AxioVision, Coot, Image Lab, MetaMorph, ImageQuant, SADABS, Imaris, DIAMOND, Gaussian, PHENIX, pClamp, GeneSpring GX, BioRender, WinGX, Pathway Analysis, Galaxy, StatView, publCIF, SDS, REFMAC5, survival, Phaser, CCP4, XDS, Clampfit, Volocity, CellQuest Pro, Living Image, Artemis, CrysAlis PRO, Mercury, Eclipse, CrystalClear, CellProfiler, Andromeda, Enrichr, GenePix Pro, Image Studio, ModFit, DIVA, Partek Genomics Suite, ModFit LT, WinNonlin, FCS Express, DALI, Scion Image, HKL2000, Multi Gauge, SAINT-Plus, CalcuSyn, PISA, Image Studio Lite, WebGestalt, rms, X-SEED, InStat, Kaluza, COLLECT, Bio-Plex Manager, SCALA, EZR, DSSP, MOLREP, ActiLife, Odyssey, Clampex, CrystalStructure, DENZO, MetaCore, WinMDI, NMRPipe

\item[Computing \& Statistics (133)] SPSS, R, ImageJ, Excel, Stata, MATLAB, Adobe Photoshop, Statistica, Python, Image-Pro Plus, Fiji, JMP, NVivo, Review Manager, ArcGIS, RStudio, Origin, lme4, SPM8, R package, G*Power, LabVIEW, SAS, Epi Info, Adobe Illustrator, REDCap, OriginPro, Mplus, EndNote, Minitab, FSL, Stata/SE, Systat, COMSOL, scikit-learn, MySQL, FreeSurfer, NIS-Elements, E-Prime, EpiData, nlme, Leica Application Suite, Qualtrics, ATLAS.ti, TensorFlow, stats, QGIS, AMOS, SciPy, randomForest, JMP Pro, MASS, SurveyMonkey, EEGLAB, ArcMap, Java, Psychophysics Toolbox, Ensemble, NumPy, Mathematica, ape, Igor Pro, Matplotlib, Weka, LIBSVM, MAS, pROC, Skype, glmnet, ANSYS, Comprehensive Meta-Analysis, FMRIB, Stata/IC, ABAQUS, Keras, lmerTest, igraph, car, REST, Amira, Spike2, SQL, SVM, MAXQDA, Design Expert, multcomp, MuMIn, JASP, PASS, Mimics, OpenCV, PyTorch, GIS, WinBUGS, Presentation, XGBoost, MATLAB script, mgcv, Stata/MP, Ethovision, LabChart, Gephi, AFNI, metafor, OsiriX, viewer, Pandas, Epi, RED, PowerPoint, NETWORK, DARTEL, SolidWorks, dplyr, Ubuntu, SoftMax Pro, CUDA, Zoom, Google Forms, SAS/STAT, Covidence, Microsoft Access, psych, tidyverse, GPower, FieldTrip, caret, Adam, NRecon, Hmisc, cellSens, Gen5, KaleidaGraph

\item[DNA (48)] Trimmomatic, ClustalX, PLINK, GATK, Clustal Omega, STRUCTURE, SignalP, UCSC genome browser, BEDTools, Integrative Genomics Viewer, SWISS-MODEL, GeneMapper, Primer Premier, Picard, PSI-BLAST, BLAT, DNAMAN, MACS, BioNumerics, Phyre2, cBioPortal, Mfold, ORF Finder, EMBOSS, CD-HIT, ProtParam, Scaffold, BioMart, ResFinder, SOAP, CIBERSORT, VarScan, BEAGLE, Burrows, Aligner, ADMIXTURE, PHASE, TransDecoder, CheckM, HaplotypeCaller, Prokaryotic Genome Annotation Pipeline, GeneDoc, XCMS, BestKeeper, MuTect, Chromas, Primer3Plus, topGO

\item[RNA \& Protein (52)] Ensembl, Cytoscape, DESeq2, DAVID, GSEA, limma, Bioconductor, TargetScan, STAR, Cluster, PolyPhen, SIFT, Haploview, MaxQuant, miRanda, Proteome Discoverer, HISAT2, PANTHER, Feature Extraction, MetaboAnalyst, WGCNA, TreeView, ANNOVAR, MutationTaster, featureCounts, GenomeStudio, SEQUEST, GEPIA, Reactome, Perseus, PicTar, affy, ClueGO, Seurat, MCODE, HOMER, survminer, Metascape, miRWalk, IMPUTE2, GeneMANIA, Venny, HTSeq-count, CompuSyn, PROVEAN, RNAhybrid, bcl2fastq, BiNGO, DESeq R, CADD, GEO2R, ComBat

\item[Microbiome \& Ecology (30)] ggplot2, QIIME, vegan, Primer3, Mothur, USEARCH, UCLUST, PAST, XLSTAT, LEfSe, FLASH, UCHIME, UPARSE, phyloseq, PICRUSt, DADA2, MaxEnt, GenStat, CAP3, CANOCO, NMDS, VSEARCH, UPGMA, MEGAN, PyNAST, STAMP, corrplot, SIMPER, MG-RAST, ade4

\item[Bioinformatics Utility (53)] BLAST, ClustalW, BLASTN, Bowtie, SAMtools, FastQC, BLASTX, TopHat, edgeR, PhyML, Blast2GO, Cufflinks, HMMER, CLC Genomics Workbench, SMART, Trinity, Cutadapt, InterProScan, TBLASTN, TMHMM, RSEM, HTSeq, pheatmap, Cuffdiff, iTOL, Jalview, BUSCO, KOBAS, SNAP, VCFtools, StringTie, TASSEL, bcftools, CASAVA, TBLASTX, RNAfold, FASTX-Toolkit, Trim Galore, megablast, MapMan, T-Coffee, PSIPRED, REVIGO, MegAlign, MapChart, JoinMap, WEGO, FreeBayes, Genomics Workbench, AgriGO, PSORT, miRDeep2, FASTX

\end{description}
\normalsize

\section{HHI trajectories by division}\label{app:hhi}
\begin{figure}[!htbp]
\centering
\includegraphics[width=0.9\textwidth]{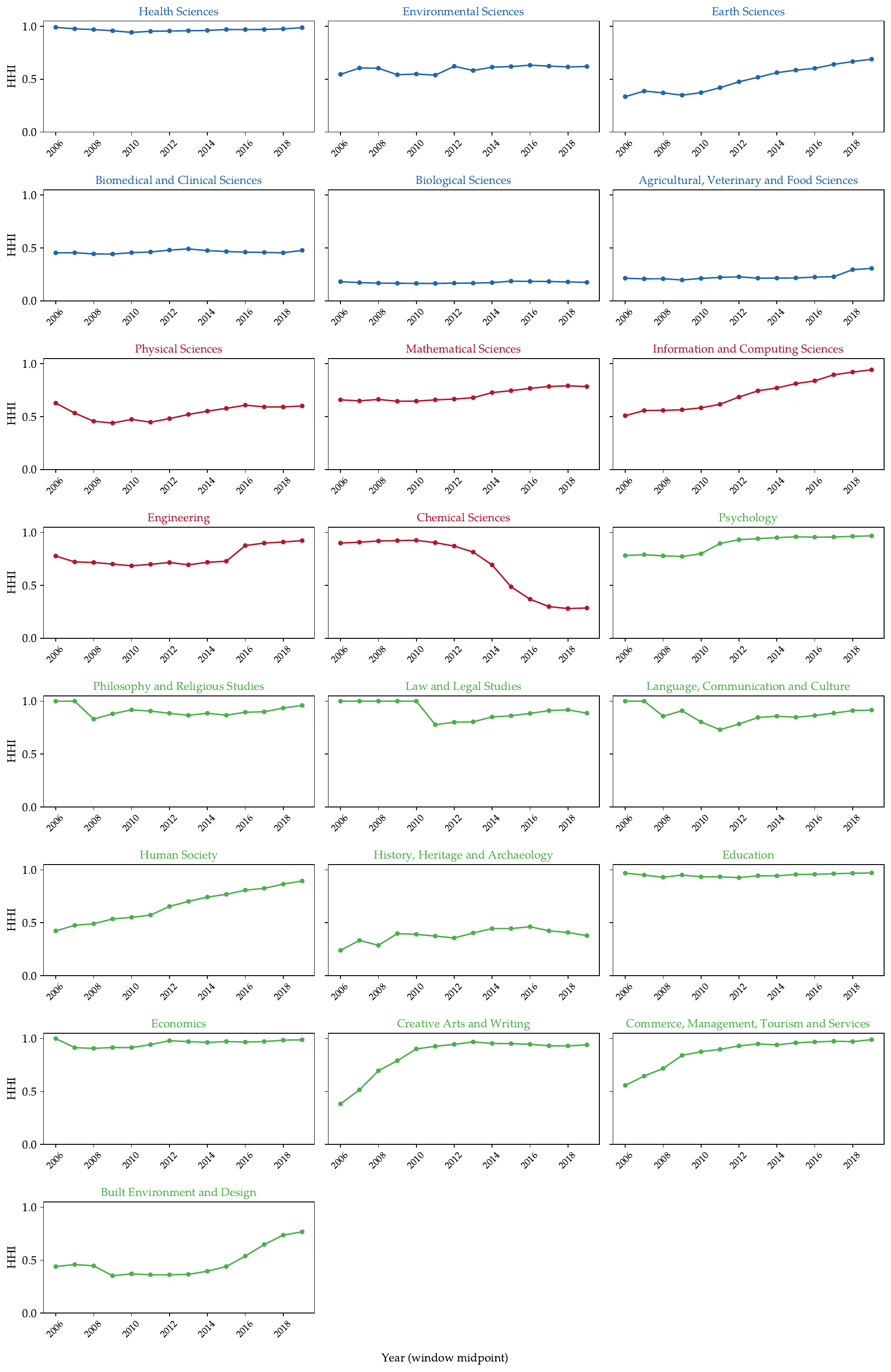}
\caption{\textbf{Most divisions show rising software community concentration.} Trajectories of software community concentration (HHI) for each of the 22 disciplinary divisions, computed over rolling 5-year windows from 2004--2008 to 2017--2021. Colors indicate broad category membership: blue for Natural and Health Sciences, red for Physical and Technical Sciences, green for Social Sciences and Humanities. Chemical Sciences is a notable outlier, with HHI declining sharply due to a compositional shift within the division. }\label{fig:hhi_appendix}
\end{figure}

\section{Jaccard stability by division}\label{app:jaccard}
\begin{figure}[!htbp]
\centering
\includegraphics[width=0.9\textwidth]{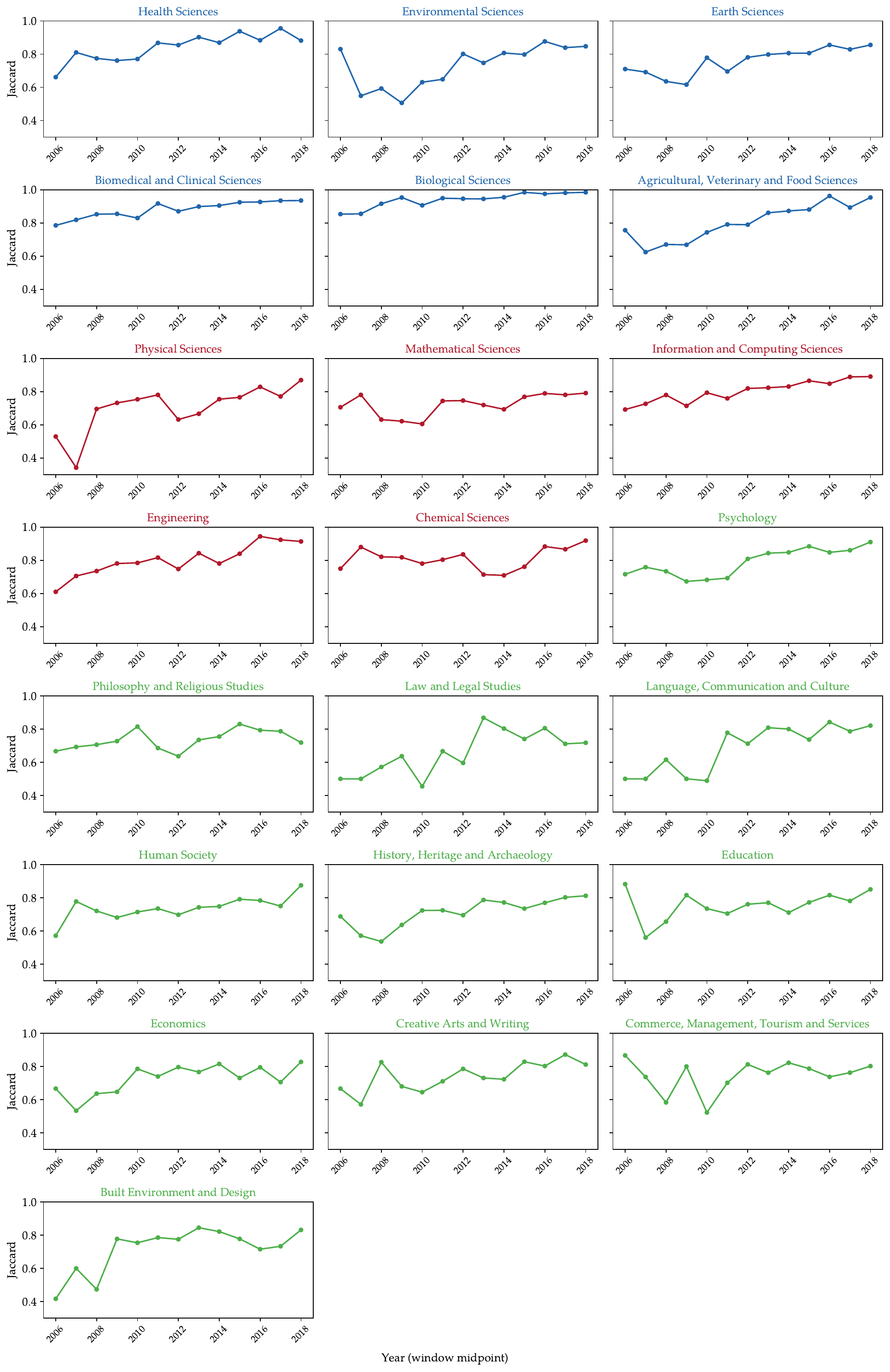}
\caption{\textbf{Most divisions show increasing stability in their software portfolios.} Stability of disciplinary software portfolios for each of the 22 divisions, measured as Jaccard similarity between consecutive rolling 5-year windows. Colors as in Fig.~\ref{fig:hhi_appendix}.}\label{fig:jaccard_appendix}
\end{figure}
\end{appendices}

\end{document}